\documentclass[preprint,prd,showpacs,showkeys,byrevtex,footinbib,eqsecnum,unsortedaddress]{revtex4}
\usepackage{amsfonts}
\usepackage{amsmath}
\usepackage{amssymb}

\usepackage{amsmath}

\usepackage{amssymb}
\usepackage{graphicx}
\usepackage{multirow}
 \usepackage{mathrsfs}
\usepackage{tikz}
\usepackage{etoolbox}
\newlength\szg
\newcommand\circled[1]{%
\settoheight\szg{#1}%
\tikz[baseline]{\pgfmathparse{
ifthenelse(#1 < 10, 1, ifthenelse(#1 < 100, 0.75, 0.5))
}
\let\hfs\pgfmathresult
\node at (0,\szg/2) {\makebox[0em][c]{\scalebox{\hfs}[1]{#1}}};
\draw (0,\szg/2) circle (\szg/2+0.5ex);
}}

\input{tcilatex}

\begin{document}

\title{A reconstruction of Florida Traffic Flow During Hurricane Irma (2017)}
\author{Kairui Feng}
\email{kairuif@princeton.edu}
\affiliation{Department of Civil and Environmental Engineering,Princeton University}

\author{Ning Lin}
\email{nlin@princeton.edu}
\affiliation{Department of Civil and Environmental Engineering,Princeton University}

\begin{abstract}
Recent Hurricane Irma (2017) created the most extensive scale of evacuation in Florida's history, involving about 6.5 million people on mandatory evacuation order and 4 million evacuation vehicles. To understand the hurricane evacuation process, the spatial and temporal evolution of the traffic flow is a critical piece of information, but it is usually not fully observed. Based on the game theory, this paper employs the available traffic observation on main highways (20 cameras; including parts of Route 1, 27 and I-75, 95, 4,10) during Irma to reconstruct the traffic flow in Florida during Irma. The model is validated with self-reported twitters. The traffic reconstruction estimates the traffic demand (about 4 million cars in total) and the temporal and spatial distribution of congestion during the evacuation. The results compare well with available information from news reports and Twitter records. The reconstructed data can be used to analyze hurricane evacuation decisions and travel behavior. 
\end{abstract}

\date{\today}
\maketitle

\section{Introduction}

Recent Hurricane Irma (2017) created an enormous scale of evacuation in Florida's history. More than 6.5 million Floridians were ordered to leave their homes. Mandatory evacuations were ordered first in the Florida Keys. As the storm approached, emergency managers in nearly every coastal county followed to issue evacuation orders \cite{FDOT2017}. As millions evacuated, heavy traffic clogged major Florida highways for five days \cite{Ralph2017}. 
To understand the hurricane evacuation process, complete and accurate dataset showing when and how many people left a given place through which road is vitally in need. 

However, such a dataset is usually not publicly available, as possibly it is still difficult for the government to monitor and record the traffic flow on the entire traffic network. The traffic data deficiency thus impedes evacuation studies. In this paper, based on the game theory, we develop a new method to reconstruct the evacuation traffic flow on all highways based on partial observations on main highways. We apply the analysis to Florida during Hurricane Irma, where the global reconstructed traffic pattern is validated using news reports and Twitter records. 

The reconstructed dataset can be used to study the evacuation process during Hurricane Irma. It can be used to validate the sociology models for evacuation decision making (e.g., \cite{urbina2003national}, \cite{hasan2010behavioral} and \cite{fu2004sequential}). It can also provide input for evacuation traffic modeling (e.g., \cite{yi2017optimization} and \cite{murray2013evacuation}). These studies together can help the government to make better evacuation orders in future hurricanes.

\section{Data and Traffic Reconstruction Model }

\subsection{Data}
\begin{figure}[!ht]
  \centering
  \includegraphics[width=0.9\textwidth]{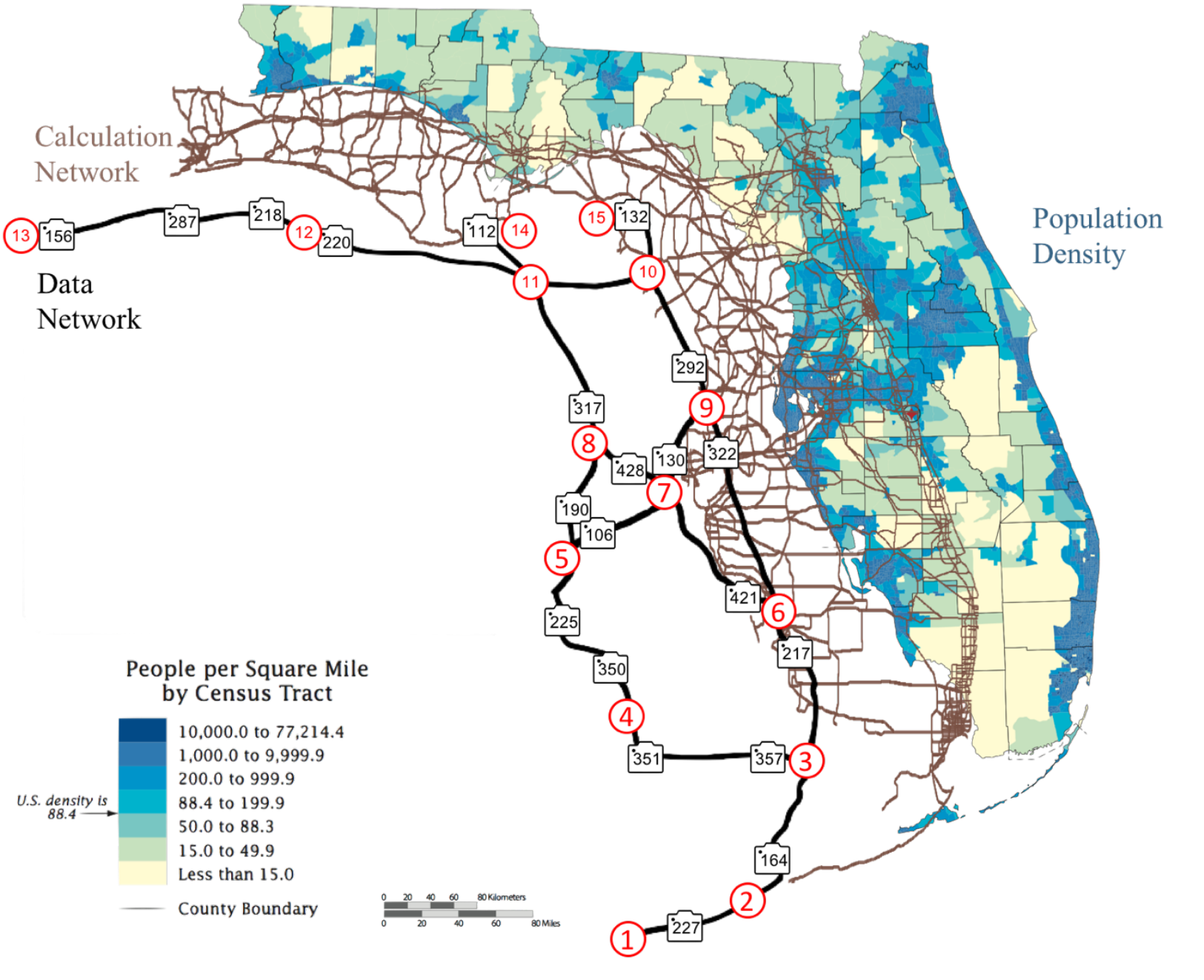}
  \caption{Overlaid maps showing data-available sparse traffic network (upper layer), calculation based detailed highway network (middle level), and population density distribution (bottom layer)}
  \label{fig:1}
\end{figure}

The traffic data for the main highways are obtained from the Florida Department of Transportation spatial analysis for Hurricane Irma \cite{FDOT2017}. The traffic data shows the number of cars passed by the cameras for every 3-hour period. The transportation network data used is a GIS database released by the Florida Department of Transportation \cite{FDOT2017}; here we consider only the roads with a speed limit over 35 mph to reduce the computational cost and maximizing the precision of the model. We also use the Census Population data \cite{Census2010} for evacuation demand estimation. The overlaying maps visualize the data in Fig.\ref{fig:1}. Since the data-available main highway network (the upper layer) is very sparse, our goal is to downscale the traffic data to the detailed traffic network (the middle layer), based on the population density distribution (the bottom layer).  In the analysis, the population for each square miles of the bottom layer is allocated to the nearest node on the traffic network.

\subsection{Reconstruction Model}
Our traffic reconstruction model is built on the most classical traffic assignment model-the (static) user equilibrium (UE) assignment model \cite{fisk1980some}. The UE model is based on Wardrop's first principle, which states that no driver can unilaterally reduce his/her travel cost by shifting to another route \cite{wardrop1952some}. The UE model is usually used to estimate the traffic flow on each road based on traffic demand (e.g., \cite{friesz1993variational}), through solving the following optimization problem \cite{leblanc1975efficient}:

  \begin{align}
    &\min_{\vec{x}} Z(\vec{x}) = \sum_a \int_0^{x_a} t_a(w)dw
   \end{align}
with the following constraint:
  \begin{align}
    &\sum_{k\in \psi_{rs}} f_k^{rs} = D_{rs},\forall\, r\in R,\forall\, s\in S\\
    &f_k^{rs} \geq 0,\forall\, k\in \psi_{rs} ,\forall\, r\in R,\forall\, s\in S\\
    &x_a = \sum_r \sum_s \sum_k f_k^{rs} \delta_{a,k}^{rs},\forall\, r\in R,\forall\, s\in S
   \end{align}

where $x_a$ is the traffic flow volume on road $a$; $\vec{x}$ is the set of the traffic flow volume on all roads on the network;  $t_a$ is the travel time on road $a$, which is a road traffic impendence function (i.e., a monotonous relationship between road traffic time and traffic amount); $D_{rs}$  is the traffic demand between origin node $r$ and destination node $s$; $f_k^{rs}$is the traffic flow volume of the $k^{\text{th}}$ route between $r$ and $s$;$\psi_{rs}$ is the set of all the routes between $r$ and $s$; $R$ and $S$ are the sets of all the origins and destinations in the traffic network, respectively; $\delta_{a,k}^{rs}$ is a Boolean function showing whether the $k^{\text{th}}$ route between $r$ and $s$ passes road $a$. The objective function defined by Eq.(1) minimizes the sum over all roads of the integral of the time cost of travel subjected to the flow conservation conditions (Eqs. 2 and 4), which requires that all the traffic demands are assigned within the traffic network. The constrain in Eq.(3) ensures that the traffic flow on each route is non-negative. 

The UE model has also been widely used to reconstruct the traffic demand\cite{van1980most}. The basic idea of traffic reconstruction is that different traffic demands (represented by a OD matrix) correspond to different traffic flow simulation results in the UE model and by minimizing the difference between the simulated results and the observed traffic flow, the OD matrix with the maximum likelihood can be found. Thus, traffic reconstruction is usually raised as an inverse problem of traffic assignment. \cite{lam1990accuracy} solved this inverse problem of the static user equilibrium and carefully tested the accuracy of the reconstruction approach. \cite{dixon2002real} generalized this approach to dynamic conditions. 

Although the above theory is straightforward, the OD reconstruction problem is usually difficult to solve due to the curse of dimensionality, the large conditional number when applying gradient-based-optimization methods to find the optimal OD and the undirected graph nature of the transportation network. However, these complexities can be reduced for the evacuation problem, where the OD matrix is reduced because evacuation is often directional (i.e., people head in the same direction) and only long-distance demand (e.g., people will not evacuate from Miami South to Miami North). We utilize these special properties to develop a new model for  reconstructing evacuation traffic demand and flow based on sparse transportation data, and we apply the model to the entire Florida during Hurricane Irma. To the best of our knowledge, this is the first global scale reconstruction of hurricane evacuation data, although some work has been done at regional scales (e.g., for New Orleans during Hurricane Katrina of 2005 \cite{dixit2011validation}).

Our evacuation traffic reconstruction is based on two assumptions : 1) Wardrop Equilibrium: travelers will strive to find the shortest (least resistance) path from origin to destination, and network equilibrium occurs when no traveler can decrease the travel cost by shifting to a new route \cite{wardrop1952some}, which is widely used in traffic assignment and  2) evacuators arriving in any node on the traffic network held the same possibility to continue evacuating as the local people. The second assumption is needed as, based only on the traffic flow count one can not separate the incoming flow and local vehicles from the out going traffic flow. The reconstruction model is essentially tracing every single car in the network to approximate the real traffic . 

In constructing the model, we define time analysis zones (TAZs) and merge all traffic nodes into their nearest TAZs. We consider 15 TAZs for Florida, as shown with red circles in Fig.\ref{fig:1}; these TAZs are defined to focus on population centers and also make sure between any two TAZs, there is at least one available traffic counter. For each time step in the simulation, the traffic flow into and out of each TAZ is recorded, the evacuation rate for each zone is estimated, and the updated population at each TAZ is calculated as the sum of the remaining local population and incoming evacuators. We use the main highway traffic data as a benchmark to adjust the simulated traffic amount on each highway, where the effect of traffic dynamics is neglected. In this case, the effect is negligible because the data comes every 3 hours, which is usually longer than the clearance time between two TAZs, consistent with the static equilibrium assumption of the UE model. For the limited cases where the clearance time is longer than 3 hours, the 3-hour time window is extended to ensure that the traffic time between the TAZs is lower than the clearance time. 

Given the defined TAZs, the origin TAZs and destination TAZs (OD pairs) can be identified. The number of OD pairs is smaller in evacuation cases than normal traffic cases because, under evacuation, the mainstream traffic goes in one direction (e.g., northwards from south Florida during Hurricane Irma). Also, the OD pairs can be classified into two groups: ODs between nearby TAZs on the top layer main highway network ("nearby") and ODs between TAZs on the second layer highway network ("long distance"). For the defined TAZs in Florida, these OD pairs are:

 Nearby: $\circled{1}\rightarrow \circled{2},\circled{2}\rightarrow\circled{3},\circled{3}\rightarrow\circled{4},\circled{3}\rightarrow\circled{6},\circled{6}\rightarrow\circled{7},\circled{6}\rightarrow\circled{9},\circled{7}\rightarrow\circled{9},\circled{5}\rightarrow\circled{7},\circled{5}\rightarrow\circled{8},\circled{7}\rightarrow\circled{8},\circled{8}\rightarrow\circled{11},\circled{11}\rightarrow\circled{14},\circled{10}\rightarrow\circled{15},\circled{11}\rightarrow\circled{12},\circled{12}\rightarrow\circled{13}$

Long Distance:$\circled{3}\rightarrow\circled{5},\circled{3}\rightarrow\circled{7},\circled{7}\rightarrow\circled{11},\circled{7}\rightarrow\circled{10},\circled{7}\rightarrow\circled{14},\circled{7}\rightarrow\circled{15}$

To analyze the entire traffic flow, one needs to consider only the direct traffic demand between TAZs (i.e., for all the OD pairs listed above). The indirect demand is always a combination of direct nearby and long distance demands. For example, if a person goes from Miami \circled{3} to Gainesville \circled{8}, the traffic demand will be calculated indirectly as from $\circled{3}\rightarrow \circled{7} \rightarrow \circled{8}$ or  $\circled{3}\rightarrow \circled{4} \rightarrow\circled{5} \rightarrow \circled{8}$ or $\circled{3}\rightarrow \circled{6} \rightarrow\circled{7} \rightarrow \circled{8}$ instead of directly from $\circled{3}$ to $\circled{8}$, because there is no direct path from $\circled{3}$ to $\circled{8}$.    

The reconstruction process follows three procedures at each time step: a) Nearby traffic analysis for the traffic flow between nearby TAZs b) Conflict analysis eliminates the traffic flow conflict when merging the nearby analysis results and c) Long distance analysis. At the end of every time step, the traffic demand happened in the last time step will be summed up and added to the population of their destination, which allows us to take into account the effect of large population evacuation (such as the $\sim$300,000 population transmission from Miami to Orlando) on the evacuation willing and pattern of local people. This method also helps to capture the global evacuation pattern more precisely.

\begin{figure}[!ht]
  \centering
  \includegraphics[width=0.9\textwidth]{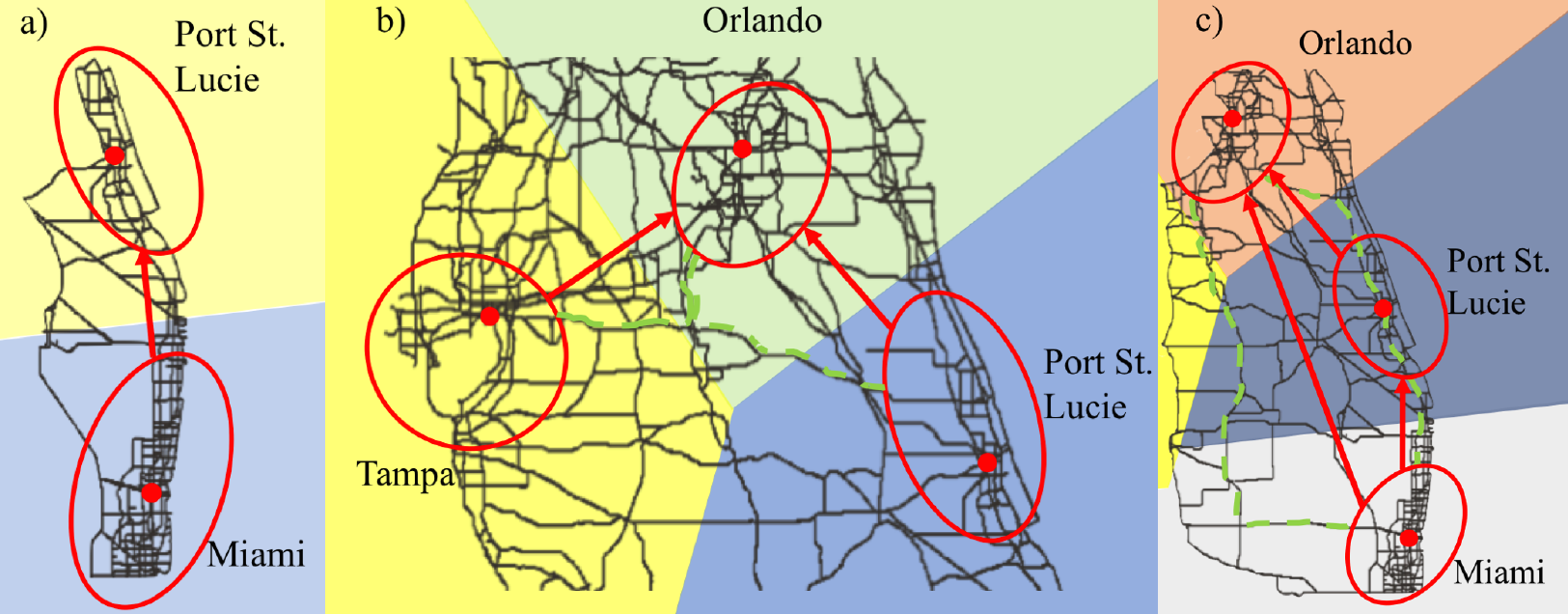}
  \caption{Examples illustrating the three procedures of traffic reconstruction: a) to analyze the traffic between two nearby TAZs; b) to analyze the traffic conflict between three traffic zones; c) to add long distance traffic demand.}
  \label{fig:2}
\end{figure}

\subsubsection*{a) Nearby Traffic Analysis}
To adapt the UE model to reconstruct the hurricane evacuation traffic demand and flow between nearby TAZs, first, we need to express the traffic demand between traffic nodes in nearby TAZs as a function of travel time. The travel time between nearby TAZs can be calculated based on the Bureau of Public Roads \cite{manual1964bureau} congestion function:
  \begin{align}
    &t_a(x_a) = t_{a0} (1+\alpha (\frac{x_a}{c_a})^\beta)
   \end{align}

where $t_{a0}$  is the free flow travel time on road $a$. $x_a$ is the volume of traffic on road $a$ (number of vehicles per hour ; vph). $c_a$ is the capacity of road a (vph). $t_a(x_a)$ is the predicted travel time on road a. Government has several approaches to increase the effectiveness of highway usage\cite{theodoulou2004alternative}. In this paper, the highway capacities are chosen following highway capacity manual \cite{manual2000highway} and the road shoulder usage during the evacuation (after Sep.$8^{\text{th}}$) is also considered following the FDOT (2017) instruction.  The parameters $\alpha$ and $\beta$ in the BPR function have large uncertainties. Zhao and Kockelman \cite{zhao2002propagation}, based on empirical data, suggests to set  $\alpha = 0.85$ with a 90\% confidence region (0.15$\sim$4.0) and $\beta = 5.5 (1.4 \sim 11)$. In our analysis, making use of the traffic volume and traffic speed sensors data released for 8 outbound highways \cite{FDOT2017}, the $\alpha$ and $\beta$ are fitted as 0.9 and 9.5 for a typical 6 lanes (3 lanes/ direction)/70 mph speed limit interstate highway (I-75 and I-95), which match with the parameters  suggested in the governmental manual \cite{horowitz1991delay}. Thus, we follow the manual, which recommends these parameters to be set as in Table 1. 

\begin{figure}[!ht]

  \centering
  \includegraphics[width=0.9\textwidth]{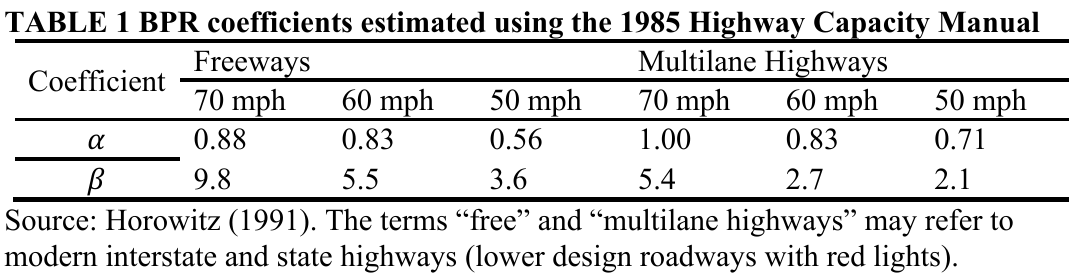}\\

\end{figure}
The traffic demand is directional in evacuation. We assume the traffic demand between a traffic node in an origin TAZ and a traffic node in a destination TAZ to be:

  \begin{align}
    &D_{rs}(\tau) =
\begin{cases}
P^{OD}_r(\tau)H_r(\tau)\frac{H_s(0)}{\sum_{\mathcal{S}\in TAZ_D}H_\mathcal{S}(0)}& \text{If} r\in TAZ_O\, \&\, s\in TAZ_D\\
0& \text{If not}
\end{cases}   \end{align}

where $TAZ_O$  represents the set of nodes in the origin  TAZ,O  and $TAZ_D$the set of nodes in the destination TAZ ,D, respectively. $P^{OD}_i(\tau)$  (\%/hour)  is the evacuation rate for node $i$ at time $\tau$ given the destination TAZ, D. $H_i(\tau)$ is the total household number at node $i$ at time $\tau$ , which is estimated as the population at each node multiply by the vehicle ownership rate. We add half of estimated tourism population for Key West because the tourism there is comparable to the population there (two people share one rental car) \cite{Census2010}. Under this setting, the traffic flow between two nearby TAZs is calculated based on the evacuating car number and its destination distribution ($\frac{H_s(0)}{\sum_{\mathcal{S}\in TAZ_D}H_\mathcal{S}(0)}$) - which follows the tradition setting of gravity model: larger population, larger attraction \cite{erlander1990gravity}. Given the population, the evacuation rate essentially determines the traffic condition, and based on Eq.5, it is a monotonously increasing function of travel time. Thus, $P^{OD}_i(\tau)$ can be estimated through minimizing the travel time.

Second, we need to incorporate the available data of traffic flows on the main highways between nearby TAZs. Noticing that the main highways are the shortest distances between nearby TAZs in daily traffic, we use the traffic cost through the main highway between two centroid nodes ($r_O\in TAZ_O, s_D\in TAZ_D$) in the two analyzing TAZs as a benchmark for the traffic assignment (see the example in Fig. 2a). 

Putting together, the evacuation reconstruction model can be expressed mathematically as:
  \begin{align}
&\min_{\vec{x},P^{OD}_r(\tau) \in [0,1]} Z(\vec{x}) = \sum_a \int_0^{x_a} t_a(w)dw
   \end{align}
with the following constraint:
  \begin{align}
&\sum_{k\in \psi_{rs}} f_k^{rs} = D_{rs} (P_r^{OD}(\tau)), \forall r\in R, \forall s \in S.\\
  &f_k^{rs} \geq 0,\forall\, k\in \psi_{rs} ,\forall\, r\in R,\forall\, s\in S\\
    &x_a = \sum_r \sum_s \sum_k f_k^{rs} \delta_{a,k}^{rs},\forall\, r\in R,\forall\, s\in S\\
 & \sum_{m \in \mathcal{H}} t_m(x_m) = \sum_{m \in \mathcal{H}} t_m(x_o^\tau) 
   \end{align}

in which $\mathcal{H}$  is the set of the main highway sections, and $x_o^\tau$  is the observed traffic flow at time $\tau$  (assumed to be uniform along the main highway). The constrain in Eq.11 ensures that the simulated traffic time over the main highway between two TAZs approaches the estimated travel time based on observed traffic when minimizing the global travel time (Eq.7). 

This evacuation reconstruction model is essentially an adjusted static UE model. The static UE assumption requires all the demands to be cleared in a given time period (otherwise dynamic UE is needed). So when the clearance time (main highway traffic time) for the system is over 3 hours (our data frequency), we extend the time window of UE symmetrically forward and backward to match the clearance time and use the traffic simulation results under the extended window to estimate the traffic demand for the original time window. The reconstruction analysis can be performed using python 2.7 for data processing and open source QGIS package AequilibraE for network equilibrium calculation \cite{camargo2015aequilibrae}.

\subsubsection*{b) Conflict Analysis}
In Part a), we reconstruct the traffic between every nearby TAZ pair. We can then merge the traffic demand on each road by linear combination. However, there is a substantial probability that the traffic flows simulated in part a) will conflict with each other. As shown in Fig 2b), the households go out from Port St Lucie $\circled{6}$ to Orlando $\circled{7}$ will conflict with households from Tampa $\circled{5}$ to Orlando $\circled{7}$ by choosing the green route, which increases the estimated traffic time in the green route, invaliding the static equilibrium established in part a). So some traffic demand should be reduced to relief this kind of conflict. Here we apply a random selection method to eliminate the excess traffic demand. For those conflicted road sections, we calculate the OD distribution of the cars on that road and also the route choice distribution following Feng et al \cite{Feng2018Post}. The probability of an individual car on road a having a specific OD, r-s, is estimated as:
  \begin{align}
	&\Pi_{a,rs} = \frac{\sum_{k\in\psi_{rs}}f^{rs}_k \delta^{rs}_{a,k}}{x_a}
   \end{align}
The probability of an individual car with an assigned OD, r-s, having a specific route, rt, is estimated as:
  \begin{align}
	&\pi_{a,rt}^{rs} = \frac{f^{rs}_{rt} \delta^{rs}_{a,rt}}{\sum_{k\in\psi_{rs}} f_k^{rs}\delta^{rs}_{a,k}}
   \end{align}

Then, for every conflicted road section, we randomly select a car on the section and assign it an O-D pair and a route based on Eqs. 12 and 13, and we calculate the travel time of the car by taking that route. If the travel time is longer than the main highway traffic, we remove the car from the demand matrix $D_{rs}$ and the route-based traffic flow $f_{rt}^{rs}$. Next, we recalculate the time cost of the route. Then we randomly select another road section from the conflicting route and repeat the analysis until there is no conflict (i.e., getting back to static equilibrium). This method is usually used directly in traffic assignment as stochastic shortest path assignment \cite{daganzo1977stochastic} and now we use that inversely to eliminate the conflict. This method considers both the population of the origin and the possibility for a person from a given origin to take a specific route, so it is fair for the whole system and can reach equilibrium approximately. Once the iteration is finished, we can recalculate the $P_r^{OD}(\tau)$  for every OD pairs to obtain the adjusted evacuation rate.

\subsubsection*{c) Long-Distance Traffic Compensation}
An example of long-distance evacuation is shown in Fig. 2c. People may rather directly go to Orlando from Miami using the lower green line than go through Port St. Lucie by the main highway. This long-distance traffic demand is not considered in the nearby evacuation analysis discussed in Parts a and b above. Here we apply a simple compensation method to incorporate this long-distance traffic demand into the data reconstruction process. We gradually add traffic demand onto the long-distance direct route until the travel time for the long distance evacuation is no longer shorter than that for the indirect routes. Then we assign the direct long-distance traffic demand into the road network and apply the random selection method in Part b) again to eliminate possibly induced route conflict. After this step, we count all the traffic flow heading out of each TAZ and recalculate the adjusted evacuation rate $P_r^{OD}(\tau)$.

		In reality, when the distant traffic demand is much lower than the nearby traffic demand, the direct long-distance evacuation may take a shorter travel time than the indirect nearby evacuation through main highways, which conflicts our assumption that the travel times are equal. Thus, the simple compensation method assuming equilibrium cannot guarantee an unbiased estimation for long distance traffic demand. Nevertheless, news reports for Hurricane Irma \cite{Osowski2017,Yanofsky2017} show that the long-distance traffic demand was comparable to the nearby traffic demand. 
		
\subsection{Results and Validation}
\begin{figure}[!ht]
  \centering
  \includegraphics[width=0.9\textwidth]{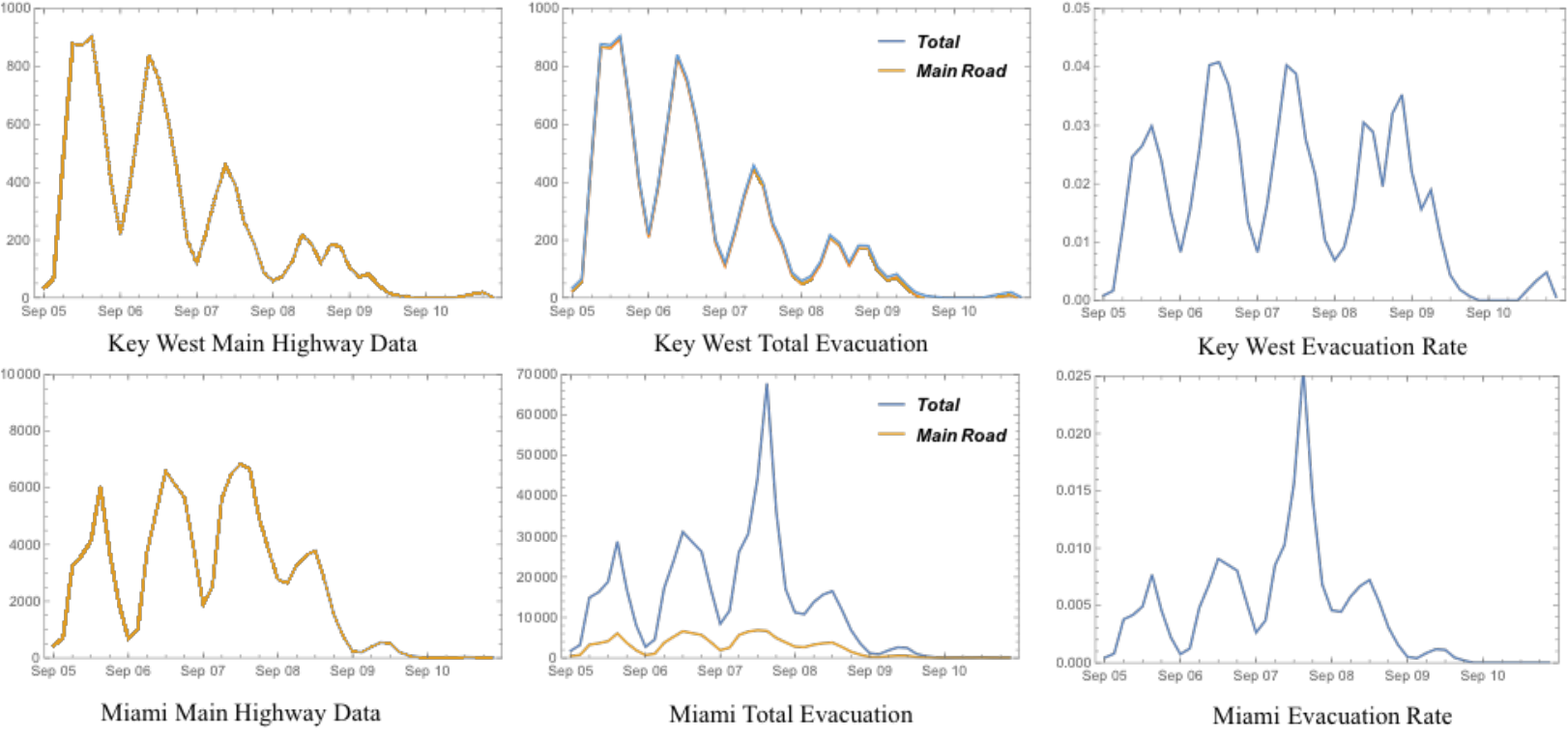}
  \caption{The evacuation traffic (vph; middle panels) and rate (\%/h; right panels) reconstructed for Key West  and Miami. The observed traffic on the main highway is shown in orange in the middle panels and also in the left panels.}
  \label{fig:3}
\end{figure}

Applying the analysis discussed above, we obtain the evacuation traffic amount for all the defined TAZs and also the traffic volume time series for all highway road sections for Florida during Hurricane Irma. Here we show the results for Key West and Miami, two most focused areas (Fig.~\ref{fig:3}).

The estimated total evacuation demand for Key West is almost the same as the observed traffic on the main highway, which is reasonable because there is only one highway from Key West Region \circled{1} to Key Largo and Homestead \circled{2} and it always takes less than 3 hours. This consistency indicates that our reconstruction model is accurate. The estimated evacuation rate changes over time for Key West and holds apparent peak. The evacuation order is released in the afternoon of Sep.$5^{\text{th}}$ and the evacuation rate keeps steady over time (with fluctuations due to day-night differences) until Sep.$9^{\text{th}}$, 12 hours before the landfall of the hurricane. The estimated total evacuation rate of Key West Region is as high as 90.5\%, indicating the significant impact of the responsible warnings released by local police officers. This large evacuation number matches with the information obtained by the authors during an impact survey for Key West soon after Hurricane Irma \cite{xian2018}. Also, according to news reports, Irma's evacuation is considered ``the largest and possibly the most successful mass evacuations in state history''\cite{Hughe2018}. The result of evacuation for the Florida Keys under Irma differs significantly from the last massive hurricane evacuation in Florida, caused by Hurricane Wilma (2005). At that time, a mandatory evacuation of residents was ordered for the Florida Keys in Monroe County. However, media reports pointed out that as many as 80\% of the residents may have ignored the evacuation order. The improved massive evacuation in Irma may be attributed to the effort of the government and media, the shocking outcome of Hurricane Harvey in Texas 7 days ago, and improved hurricane awareness of Floridians. 

The calculated evacuation demand of Miami changes dramatically between 4 $\sim$ 7 times of the main highway traffic. This large difference is due to the extending-in-all-directions traffic network in Miami and also the multiple destinations such as Tampa and Orlando. The evacuation rate of Miami was much lower than the evacuation rate of Key West, as the risk for Miami was much lower than for Key West, which was supposed to face a direct landfall. And plenty of airlines and shelters in Miami also helped local people to leave through other means than evacuation on the traffic network. The low gasoline availability (40\% shortage reported) during that period may have restrained the exodus willing for Miami people on Sep.$8^{\text{th}}$ \cite{Egan2018}. The peak traffic happened at noon of Sep.$7^{\text{th}}$, although the evacuation order was released at noon on Sep.$6^{\text{th}}$. This delay in the traffic peak reflects the fact that people often need 6 to 12 hours for preparation and tend to avoid evacuation at night \cite{lindell2005household}. In total, $\sim$1,870,000 cars evacuated out of Miami and supposing there are 4,000,000 cars in the Miami region based on the census data, about 46\% of Miami people went on evacuation.

The reconstructed traffic data can also be used for estimating the fatality rate. Irma caused 10 direct deaths and an additional 82 indirect deaths as a result of its strong winds, heavy rains, and high surf across the southeastern United States\cite{knabb2005tropical}. As reported, Hurricane Irma led to 14 victims (15\% of total) in Key West. Considering that only about 2000 households remain without evacuation, the fatality rate for people left in the Florida Keys is as high as about 0.7\%. In Miami-Dade County, no fatality is reported. The quantification of hurricane fatality risk can support fatality-control based evacuation policy design.

It is difficult to directly and fully validate our model because there is no observation other than the traffic flow on the main highways. Also, we can not do cross-validation by taking out some observed traffic data from our training set, because all the (quite limited) main highway data are necessary benchmarks for reconstruction. However, we can partially validate the model based on Twitter records, where some evacuators reported their origin, destination, and time cost. For example, one person reported that he went from Miami \circled{3} to Gainesville \circled{11} for 14 hours on Sep.$7^{\text{th}}$. Our model estimates that, if they left Miami between morning and noon on Sep.$7^{\text{th}}$, it would take them 12$\sim$14 hours to reach Gainesville. Another person at Tampa twitted that it took her 5 hours in the afternoon of Sep.$8^{\text{th}}$ to travel the distance that usually takes her 2 hours; our model estimates that the time cost for a normal 2-hour trip is between 4$\sim$10 hours on Sep.$8^{\text{th}}$ depends on which direction she was heading. The Twitter data, however, are highly unstructured because everyone report their traffic time in different formats. We have about 30 records to validate our model. In the interests of privacy, we will not show all the data here. 

\section{conclusion}
In this paper we developed an evacuation traffic reconstruction model based on the static UE assumption. We apply the model to estimate the temporal changing pattern of all road ways (speed limit $\geq$ 35 mph) in Florida during Hurricane Irma based on available main highway traffic data. The global reconstructed traffic pattern is validated with news reports and Twitter records. 
Traffic records of two main evacuating cities - Florida Keys and Miami - are analyzed. This largest evacuation in the US history may have led $\sim$46\% Miami people and $\sim$90\% Keys people to evacuate. The total traffic amount was estimated to be much higher (4$\sim$7 times) than the traffic observed on the main highway for Miami.
The reconstructed traffic dataset can be used to study evacuation decision making, travel behavior and quantify fatality rate, improve evacuation orders and governmental risk management. 

\section{Acknowledgements}
This material is based upon work supported by the National Science Foundation (grant 1652448). We also thank the Florida Department of Transportation for providing the evacuation data during Hurricane Irma.

\section{Author contribution}
K.F. and N.L. conceived of the presented idea and wrote the manuscript. K.F. developed the methodology and performed the computations.

\end{document}